\documentclass{article}

\usepackage{PRIMEarxiv}

\usepackage[utf8]{inputenc} 
\usepackage[T1]{fontenc}    
\usepackage{hyperref}       
\usepackage{url}            
\usepackage{booktabs}       
\usepackage{amsfonts}       
\usepackage{nicefrac}       
\usepackage{microtype}      
\usepackage{lipsum}
\usepackage{fancyhdr}       
\usepackage{graphicx}       
\graphicspath{{media/}}     

\title{Learning Ecology with Vera using Conceptual Models and Simulations}

\author{
  Spencer Rugaber \\
  School of Computer Science\\
  Georgia Institute of Technology\\
  \texttt{spencer@cc.gatech.edu} \\
  \And
  Scott Bunin \\
  School of Computer Science\\
  Georgia Institute of Technology\\
  \texttt{sbunin3@gatech.edu} \\
  \And
  Andrew Hornback \\
  School of Computer Science \\
  Georgia Institute of Technology\\
  \texttt{ahornback6@gatech.edu} \\
  \And
  Sungeun An \\
  School of Computer Science\\
  Georgia Institute of Technology\\
  \texttt{sungeun.an@gatech.edu} \\
  \And
  Ashok Goel \\
  School of Computer Science\\
  Georgia Institute of Technology\\
  \texttt{ashok.goel@cc.gatech.edu} \\
}

\begin{document}
\maketitle

\begin{abstract}
Conceptual modeling has been an important part of constructionist educational practices for many years, particularly in STEM (Science, Technology, Engineering and Mathematics) disciplines. What is not so common is using agent-based simulation to provide students feedback on model quality. This requires the capability of automatically compiling the concept model into its simulation. The VERA (Virtual Experimentation Research Assistant) system is a conceptual modeling tool used since 2016 to provide introductory college biology students with the capability of conceptual modeling and agent-based simulation in the ecological domain. This paper describes VERA and its approach to coupling conceptual modeling and simulation with emphasis on how a model's visual syntax is compiled into code executable on a NetLogo simulation engine. Experience with VERA in introductory biology classes at several universities and through the Smithsonian Institution's Encyclopedia of Life website is related.
\end{abstract}

\keywords{Agent-based simulation, conceptual modeling, ecology, interactive learning environment, visual language}

\section{Introduction}
VERA (Virtual Experimentation Research Assistant) is an interactive modeling tool aimed at STEM (Science, Technology, Engineering and Mathematics) undergraduate students. It is based on the hypothesis that a constructionist approach facilitates learning \cite{1_papert1991situating}, that is, students gain deep knowledge when they construct it themselves. Its core domain is the field of ecology.

Key contributions of VERA are its conceptual framework and a practical technology for compiling conceptual models into agent-based simulations. Typically, frameworks and technologies for interactive learning environments treat conceptual models and agent-based simulations separately. For example, in Betty’s Brain \cite{2_biswas2005teachable, 3_leelawong2008betty}, middle school students can either construct conceptual models or examine agent-based simulations, but the two are not directly related. In contrast, VERA automatically generates the agent-based simulations from the conceptual models: A student first interactively generates a conceptual model and then asks VERA to evaluate the model; VERA generates the simulation and displays the results to the student. This affords students with a powerful capacity to both generate and evaluate their models. This section provides the background and motivation for VERA.

\subsection{Systems Thinking}
Modern society is characterized by complex natural, social, engineering, and socio-technical systems. A system is a collection of interacting components such that the behaviors of the system arise out of interactions among the components. A complex system is a system that has a large number of components and processes at multiple levels of organization \cite{4_forrester2021principles, 5_simon2018science}. Thus, to understand and tackle problems of complex systems, individuals must employ systems thinking, a strong ability to reason about and understand complex systems in a practical and useful way. Systems thinking is characterized by thinking about relationships among components and processes, thinking at multiple levels of abstraction, and thinking about systems-as-a-whole \cite{5_simon2018science, 6_silk2008core, 7_meadows2008thinking}. Systems thinking enables learners to engage in scientific reasoning, such as understanding patterns, causes and effects, and interdependence between phenomena and explanation \cite{8_mclucas2003decision, 9_jacobson2006complex}.

\subsection{Scientific and Conceptual Modeling in Systems Thinking}

Systems thinking encompasses multiple phases including model construction, model usage, model evaluation, and model revision \cite{10_dickes2016development, 11_jordan2018developing, 12_joyner2014mila}. Modeling of complex systems enriches systems thinking by highlighting the central phenomenon of a system \cite{10_dickes2016development, 11_jordan2018developing}. In response, researchers have developed interactive modeling environments to promote understanding and reasoning about complex systems for novice learners. Using these modeling environments as instruments, researchers have found significant learning effects \cite{2_biswas2005teachable, 11_jordan2018developing, 13_bredeweg2013dynalearn, 14_wilensky2006thinking, 15_bridewell2006supporting, 16_vattam2011understanding, 17_joyner2015improving, 18_gilbert2018developing}. 

Qualitative conceptual modeling tools are designed to help learners specify system elements, their relationships and whether the elements are positively or negatively correlated. Thus, they focus on concepts and relationships at the expense of numerical precision and predictive abilities. With these tools, learners first create conceptual models of complex systems and then qualitatively simulate them \cite{3_leelawong2008betty, 13_bredeweg2013dynalearn, 19_forbus2005vmodel}. 

\subsection{System Modeling in STEM Education}

Cognitive theories of scientific discovery propose that scientific progress is achieved through the construction, evaluation, and revision of models \cite{20_clement2008creative, 21_nersessian2010creating}. When scientists encounter abnormal or atypical phenomena that cannot be explained with existing knowledge \cite{22_darden2009strategies}, they generate explanatory hypotheses. As they evaluate their models, hypotheses that are not supported by data are discarded. Novice learners also engage in constructing, revising, and updating their knowledge through modeling \cite{23_sins2005difficult, 24_schwarz2009learning, 25_white1998inquiry, 26_nersessian1999model}. The basis of such arguments has its roots in cognitive theories, including constructionism and situated cognition. In constructionism, learners actively construct their own knowledge and understanding \cite{27_papert2020mindstorms}, while situated cognition emphasizes the importance of social context and practical application \cite{28_lave1991situated, 29_lave1991situated_duplicate}. 

Both constructionism and situated learning have been influential in the development of inquiry-based modeling \cite{25_white1998inquiry}, as these approaches prioritize hands-on, experiential learning and emphasize the importance of learners actively engaging with the material. In this view, students learn by actively constructing and evaluating questions and models. Since the 1960s, there has been a great deal of research and curriculum development aimed at teaching scientific reasoning to K-12 students. Significant research has identified the value of involving students in authentic scientific processes, such as inquiry-based modeling, in early science education \cite{12_joyner2014mila, 17_joyner2015improving, 23_sins2005difficult, 25_white1998inquiry, 30_vanlehn2013model} and in undergraduate science education as well \cite{11_jordan2018developing, 18_gilbert2018developing}. 

\subsection{Teaching Ecology }

One kind of complex systems are ecologies, which study the interaction of various biotic and abiotic components and processes \cite{31_gaucherel2020understanding, 32_arnold2015definition}. Learning ecology through modeling is realized by interactive modeling environments enabling students to construct models by selecting, placing and connecting icons on a drawing canvas. VERA, the system described here, is one such conceptual modeling tool. 

Agent-based simulations (ABSs) are especially appropriate for  evaluating ecologies. However, many students, including college students, are not skilled at building agent-based simulations because of lack of expertise in mathematics and programming. In contrast, these students are often proficient at constructing conceptual models. A key issue in teaching about ecological systems thus becomes how to provide students feedback on the conceptual models they create. VERA is an artificial intelligence (AI) assistant that compiles conceptual models into ABSs useful for evaluating the students' hypotheses. 

\begin{figure}[h]
  \centering
  \includegraphics[scale=0.75]{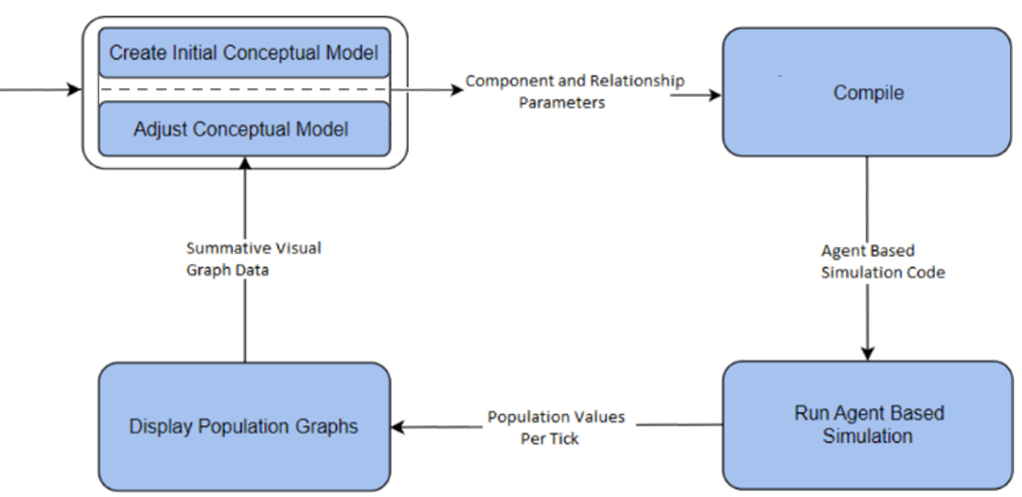}
  \caption{VERA Information Flow}
  \label{fig:fig1}
\end{figure}
\section{Running Example}
VERA is an interactive modeling tool that enables students to better understand ecological systems. As shown in Figure \ref{fig:fig1}, after drawing and parameterizing a conceptual model (upper left), users can request the model be compiled into a program for execution by a NetLogo simulator (upper right). As the simulation runs (lower right), a graph showing population levels over time is updated (lower left). After evaluating the results, the user can iterate the process by refining the parameters of the model and rerunning the simulation. This section presents a running example of how a simple ecological model is processed by VERA. 

\begin{figure}[h]
  \centering
  \includegraphics[scale=0.75]{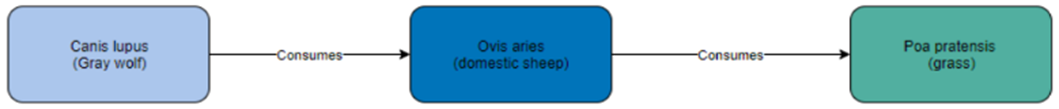}
  \caption{VERA Model for a Predator–Prey Ecology}
  \label{fig:fig2}
\end{figure}

\begin{figure}[h]
  \centering
  \includegraphics[width=1.0\textwidth]{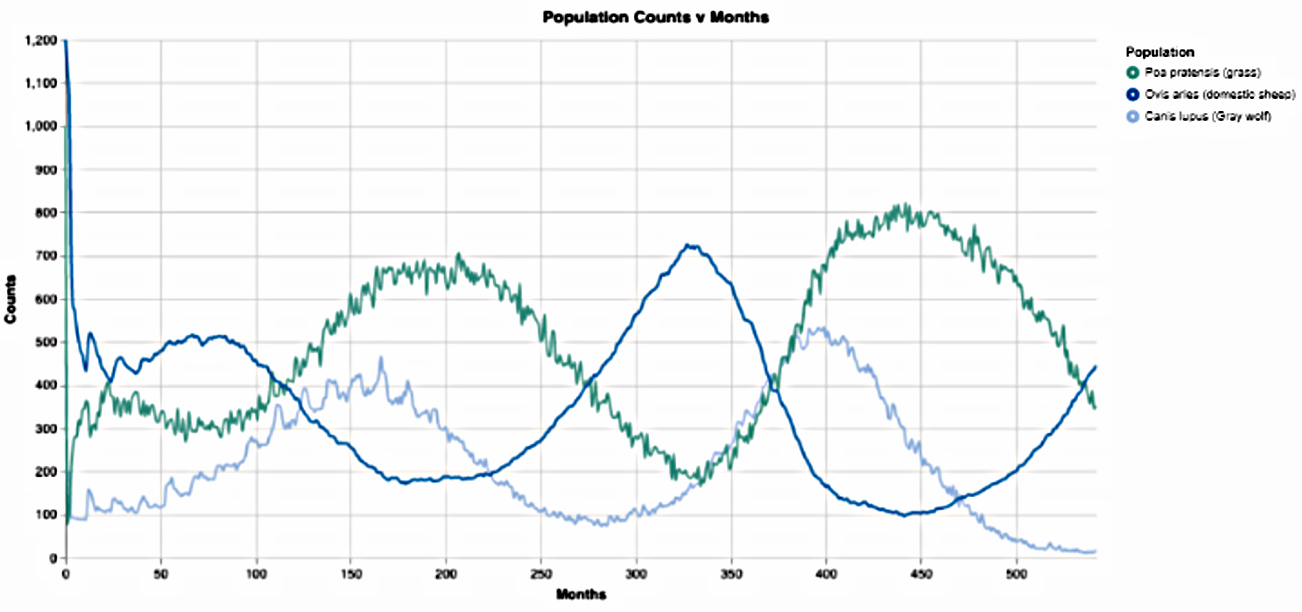}
  \caption{Wolf–Sheep–Grass Simulation Output}
  \label{fig:fig3}
\end{figure}

Figure \ref{fig:fig2} illustrates a predator–prey Relationship. The model shows three biotic Components (colored roundtangles), denoting gray wolves (Canis Lupus), domestic sheep (Ovis aries) and Kentucky bluegrass (Poa pratensis), and two instances of the Consumes Relationships (arrows). When simulated, the sheep population increases with the availability of grass, which in turn, leads to an increase in the wolf population. A greater number of wolves, in turn, leads to a decrease in the population of sheep, which, as the sheep population diminishes, results in a reduction in the number of wolves. The cycle repeats as the consumables become more or less available as shown in Figure \ref{fig:fig3}. 

The horizontal axis of the graph conveys the passage of simulated time in months, while the vertical axis presents population levels. Three curves are plotted, one each for grass (green), sheep (blue) and wolves (light blue). 

\begin{figure}[h]
  \centering
  \includegraphics[scale=0.75]{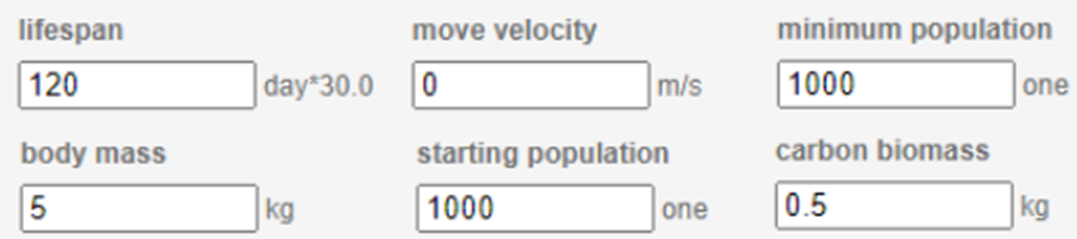}
  \caption{Parameter values for Kentucky bluegrass}
  \label{fig:fig4}
\end{figure}

VERA models include user-adjustable parameters for each Component and Relationship. Clicking on a Component or Relationship displays its available parameters. Values of the Component parameters control factors such as the initial population level and average lifespan of individuals in that Component’s population. Relationship parameters can also be modified, for example to adjust the Consumption Rate and Interaction Probability parameter values. By editing parameter values, users can refine their models and test different hypotheses about how the ecology functions. The set of parameters shown in Figure \ref{fig:fig4} is taken from the Grass biotic Component in the Predator-Prey model. 
\section{VERA's Approach}
\subsection{ACT and MILA-S}
A predecessor to VERA is ACT (Aquarium Construction Toolkit), which supports modeling physical systems with the Structure-Behavior-Function language \cite{16_vattam2011understanding}. ACT was used in three middle school classrooms in New Jersey with a total of 157 students in the 7th and 8th grade and generated positive learning outcomes. The software was used to model a classroom aquarium that learners focus in greater depth on details of underlying scientific causes of a healthy ecosystem such as the levels of ammonia, oxygen and nitrates. ACT has enabled students to learn ecological concepts by the process of constructing a model, critiquing it and then revising it. ACT and VERA both make use of NetLogo \cite{14_wilensky2006thinking}.  

Another predecessor to VERA is MILA-S (Modeling and Inquiry Learning Application for Simulation) \cite{12_joyner2014mila}. It was designed for educational use and was tested with 50 Georgia middle school students. MILA-S enables students to set parameter values and choose Relationships between Components, such as Produces and Consumes. The system was designed to enable students to model visually and to rapidly prototype and test their ideas via a NetLogo simulation. This approach proved useful to middle school students in implementing a modeling cycle comprising conceptual model design, simulation creation, evaluation of results and revisions based on model results. However, some students were distracted from the intended lesson goals of observing specific ecological concepts and this was used in VERA’s ideation in that it could be valuable to show a summarized abstraction of the simulation.    

\subsection{VERA System Architecture}

\begin{figure}[h]
  \centering
  \includegraphics[scale=0.7]{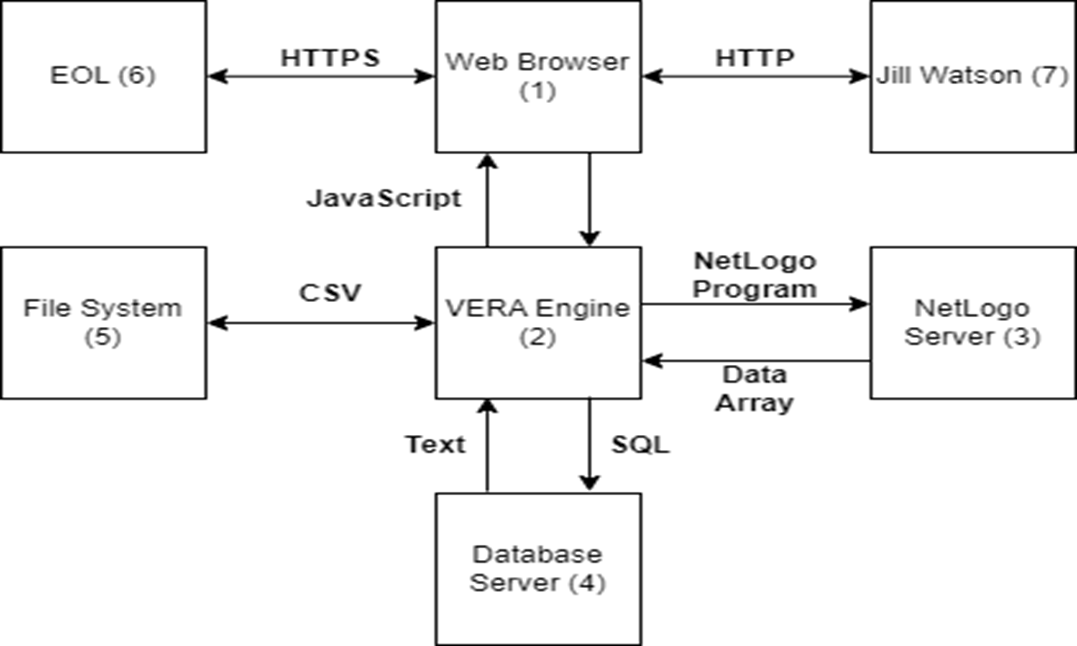}
  \caption{VERA System Architecture}
  \label{fig:fig5}
\end{figure}

The VERA system architecture is depicted in Figure \ref{fig:fig5}. Its user-facing aspect is provided by a web browser (1) executing JavaScript functions. A user constructs models, provides parameter values, either manually or automatically obtained from the Smithsonian Institution's Encyclopedia of Life (EOL) \cite{33_parr2014encyclopedia}, initiates a simulation and views the resulting graphs in the browser. Model analysis and overall process control is provided by the VERA Engine (2). The Engine stores user models in a MySQL database (4). Engine processing includes compilation, authentication, storage management and process control. 

The VERA compiler processes a parameterized user model to produce code written in a subset of the NetLogo language. The subset, described below, acts as a virtual machine for running ecological simulations. The code is sent to a headless NetLogo server (3), which runs the simulation specified in the compiled code and returns a stream of population records that are forwarded to the browser for visualization by the Vega graphics package \cite{34_satyanarayan2015reactive}. In addition to the processing described above, VERA also produces and stores log data and imports and exports CSV data describing population histories from its host machines file system (6). 

\subsection{NetLogo}

Agent-based modeling (ABM) tools allow learners to simulate the behavior and interactions of individual agents within a system, making them powerful instruments for exploring complex, dynamic phenomena across a range of domains \cite{12_joyner2014mila, 35_resnick1996beyond, 36_wilensky1999thinking}. These models typically consist of many agents, each executing simple rules or programs \cite{27_papert2020mindstorms, 35_resnick1996beyond, 37_basu2013ctsim}. By specifying the local rules that govern agent behavior, learners can investigate how complex, system-level behaviors emerge from decentralized interactions among individual agents.

Agent-based modeling is particularly useful for understanding the dynamics of decentralized and self-organizing systems, such as ant colonies, traffic flow, or ecosystems, where emergent behavior arises from simple interactions. ABM environments enable learners to observe and analyze these emergent patterns, strengthening systems thinking and intuition about non-linear processes. However, the programming requirements for constructing such models can pose a challenge for learners without prior coding experience.

One of the most widely used ABM environments in education is NetLogo \cite{14_wilensky2006thinking, 30_vanlehn2013model, 36_wilensky1999thinking}. Authored by Uri Wilensky at Northwestern University, NetLogo provides a user-friendly simulation environment for modeling interactions among multiple agents using a derivative of the LOGO educational language \cite{38_abelson1986turtle}. Learners write code to create and control agents, run simulations, and observe results through visualizations, graphs, and tables. Its intuitive interface allows users to manipulate variables and animation settings, facilitating experimentation with different scenarios and exploration of the effects of parameter changes. NetLogo is publicly available and runs on multiple operating systems. Simulations can be executed as part of a Java application, through a web interface, or “headless” via an Application Programming Interface (API), which is the mode used by VERA \cite{36_wilensky1999thinking}.

\subsection{The VERA Conceptual Modeling Language}

The VERA modeling canvas provides a visual syntax for representing ecological phenomena, including Components and the Relationships between them. VERA users construct ecological models by drawing on the canvas. The modeling process includes selecting and configuring Components, usually denoting species (biotic Components) or nonliving (abiotic Components) factors such as, for example, an oil spill. 

Components are connected by Relationships that describe interactions between the Components. Available Relationships include Consumes, Destroys, Produces, and Affects. Detailed behavior can be controlled by adjusting Component and Relationship parameter values.

\begin{figure}[h]
  \centering
  \includegraphics[width=1.0\textwidth]{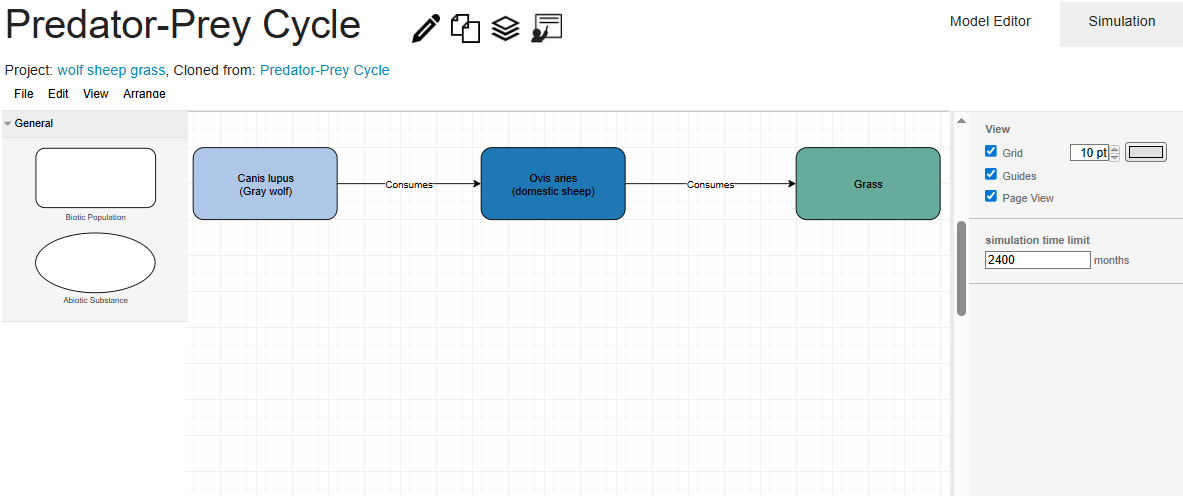}
  \caption{The VERA Conceptual Modeling Interface}
  \label{fig:fig6}
\end{figure}

The conceptual modeling interface contains options for managing models within projects, a selection of Components to add to the model and an area for editing parameter values and other settings. The options for managing the project include making copies of conceptual models and grouping multiple models under a user-named project. The wolf sheep grass project containing the running example is shown in Figure \ref{fig:fig6}. 

Component and Relationship parameter values can also be altered. The simulation produces vastly different results depending not only on how the individual Component parameter values are set, but also on how each of the Consumes Relationships are parameterized. For example, if the Reproductive Maturity for domestic sheep is late and their Reproductive Interval is large, they will take longer before they begin to reproduce and reproduce less often. Thus, even if the Interaction Probability of the wolf-to-sheep Consumes Relationship is set to a relatively high value, the gray wolf population will still go extinct because there are few domestic sheep available. Table \ref{tab:t1parameters} shows the default parameter values for the running example. 

\begin{table}[h!]
\centering
\caption{Parameter Values for the Running Example}
\label{tab:t1parameters}
\begin{tabular}{lrrr}
\hline
\textbf{Parameters} & \textbf{Wolf} & \textbf{Sheep} & \textbf{Grass} \\
\hline
Lifespan (months) & 180 & 252 & 120 \\
Body mass (kg) & 30 & 19.66 & 5 \\
Starting Population & 200 & 1200 & 1000 \\
Offspring Count & 5 & 1 & 0 \\
Reproductive Maturity (months) & 30 & 24 & 0 \\
Reproductive Interval (months) & 12 & 12 & 0 \\
Minimum Population & 0 & 0 & 1000 \\
\hline
\end{tabular}
\end{table}

Table \ref{tab:t2bio_parameters} provides a description for the thirteen biotic Component properties. Note that not all parameters are meaningful for all biotic Components. Table \ref{tab:t3abiotic_parameters} presents similar information for abiotic Components. The previously mentioned running example does not contain any abiotic Components. However, with a suitable adjustment, the grass population could be limited by an abiotic factor such as water availability. For example, the simulation might begin at 100 units of water. Then, 77\% of the water might be replenished for each timeframe, as represented by the Minimum Amount parameter. In addition to parameterized Components, VERA supports five predefined Relationships. These relationships have their own parameters, which are listed in Table \ref{tab:t4relationship_parameters}. 

\begin{table}[h!]
\centering
\caption{Parameters for Biological Components}
\label{tab:t2bio_parameters}
\begin{tabular}{lp{8cm}l}
\hline
\textbf{Parameter} & \textbf{Description} & \textbf{Unit of Measure} \\
\hline
Carbon Biomass & Average carbon biomass in an individual organism, or per m$^2$ in a density-based population & Kilogram \\
Respiratory Rate & Average basal metabolic rate, measured as respiration (loss) of carbon biomass & Kilograms per second \\
Photosynthesis Rate & Average addition of carbon biomass from photosynthesis for a square meter of density-based populations & Kilogram per second \\
Assimilation Efficiency & Efficiency of assimilating carbon biomass via consumption (0.0 - 1.0) & One Unit \\
Move Direction & General direction of organism movement measured in compass bearing degrees & Directional degree \\
Move Velocity & General velocity of organism movement & Meter per second \\
Lifespan & Average lifespan of organisms in this population & Day * 30 \\
Reproductive Maturity & Age when organisms in this population can begin reproduction & Day * 30 \\
Reproductive Interval & Frequency with which organisms in this population can reproduce & Day * 30 \\
Offspring Count & Average number of offspring per spawning individual during a reproduction cycle & One Unit \\
Starting Population & For populations with individual organisms, number of individuals to start with. For populations measured in m$^2$ coverage, the number of m$^2$ to start with & One Unit \\
Minimum Population & For populations with individual organisms, the minimum number of individuals to maintain. For populations measured in m$^2$ coverage, the number of m$^2$ to maintain at minimum & One Unit \\
Body Mass & For populations with individual organisms, average body mass per organism. For populations measured in kg/m$^2$ densities, the density value & Kilogram \\
\hline
\end{tabular}
\end{table}

\begin{table}[h!]
\centering
\caption{Parameters for Abiotic Components}
\label{tab:t3abiotic_parameters}
\begin{tabular}{lp{8cm}l}
\hline
\textbf{Parameter} & \textbf{Description} & \textbf{Unit of Measure} \\
\hline
Amount & Starting amount of the substance distributed across the ecosystem & Kilograms \\
Minimum Amount & Minimum amount of the substance to maintain in the ecosystem & Kilograms \\
Growth Rate & An abiotic substance can act to stimulate or degrade the growth of affected biotic populations & Percentage \\
\hline
\end{tabular}
\end{table}

\begin{table}[h!]
\centering
\caption{Parameters for Relationships}
\label{tab:t4relationship_parameters}
\begin{tabular}{lp{3cm}p{6cm}p{5cm}}
\hline
\textbf{Relationship} & \textbf{Parameter} & \textbf{Description} & \textbf{Example} \\
\hline
Affects & Interaction Probability & Chance of interacting (expressed as a percentage) when within range of the targeted Component & Fertilizer can be unevenly distributed (e.g., a 50\% parameter value that increases growth only half the time) \\
Affects & Growth Rate & Multiplier applied to the number of Component instances added during each time frame & Growth Rate of 0.10 increases the amount of grass by 10\% \\
Consumes & Interaction Probability & Probability (expressed as a percent) of interacting when within range of the target Component & A wolf may be within range of a sheep with Interaction Probability 0.10, indicating it is 10\% likely to consume the sheep \\
Consumes & Consumption Rate & Percent of the target Component’s carbon-biomass that is removed from the target and given to the consuming Component & With a parameter setting of 0.20, the grass has 20\% of its available carbon-biomass removed by the sheep \\
Destroys & Interaction Probability & Chance of interacting when within range of the target Component & A wolf within range of a sheep with Interaction Probability 0.10 is 10\% likely to destroy the sheep, but not consume it \\
Destroys & Destruction Rate & Percent of target removed per interaction & If the Destruction Rate has a parameter of 0.10, the amount of grass is decreased by 10\% \\
Produces & Production Rate & Creation of the target Component in kilograms per second & If the sheep produce an abiotic component with a parameter setting of 1, it will accumulate 1 abiotic unit per calculation tic \\
\hline
\end{tabular}
\end{table}

\subsection{NetLogo Simulation Engine}

Once the user has drawn and parameterized a VERA model, it can be simulated. The user initiates this process by selecting the Simulation tab, which causes the visual model to be compiled into a NetLogo program. The simulation output shows how the populations of the biotic Components change over subsequent months. The resulting population changes are then displayed graphically. 

In addition to the line graphs provided to the user for analysis, the user may also use the Export Data to CSV option to obtain the same information in tabular form. As seen in Table \ref{tab:t5_exported_data}, the values may then be used for quantitative analysis. For example, in the data shown, a user analysis can see the significance in the transition from Tick 2 to Tick 3 as the wolf population crashes. The data tables are also useful for comparing results of multiple runs. 

\begin{table}[h!]
\centering
\caption{Example of Exported Data}
\label{tab:t5_exported_data}
\begin{tabular}{lrrr}
\hline
\textbf{Month} & \textbf{Wolf} & \textbf{Sheep} & \textbf{Grass} \\
\hline
0  & 200 & 1200 & 1000 \\
1  & 200 & 1128 & 57   \\
2  & 200 & 1072 & 104  \\
3  & 85  & 733  & 171  \\
4  & 84  & 594  & 250  \\
5  & 84  & 582  & 276  \\
6  & 81  & 547  & 289  \\
7  & 80  & 517  & 306  \\
8  & 79  & 490  & 311  \\
9  & 79  & 477  & 330  \\
10 & 79  & 460  & 353  \\
11 & 79  & 446  & 324  \\
\hline
\end{tabular}
\end{table}

\subsection{NetLogo Simulation Engine–Execution Environment}
NetLogo agent-based simulations consist of agents of different breeds inhabiting geographical patches. These simulations have been used in numerous areas including art, biology, chemistry, physics, computer science, earth science, games, mathematics, networks, philosophy, psychology, social science and system dynamics \cite{44_an2022effects}. 

An ABS specifies the interaction between separate agents that follow certain rules and make decisions based on the data in their own state and from the environment. The simulations are time-based, organized by clocked ticks, rather than event based, in which the agents respond asynchronously to events. 

Ticks are used in ABSs to keep track of the current time frame. During each tick, the NetLogo code (methods) compiled for each Component and Relationship are processed. Methods can lead to location changes and parameter value updates. 

Movement of agents over patches makes use of an X-Y rectilinear coordinate system. Each agent is described in terms of its current location, heading and speed. Agent-to-agent interactions are determined based on the distance between the two agents involved and the probabilities of their interactions. 

Ticks are used in agent-based simulations to keep track of the current time frame. During each tick, the NetLogo code (methods) compiled for each Component and Relationship are processed. Methods can lead to location changes and parameter value updates. 

Movement of agents over patches makes use of an X-Y rectilinear coordinate system. Each agent is described in terms of its current location, heading and speed. Agent-to-agent interactions are determined based on the distance between the two agents involved and the probabilities of their interactions. 

\subsection{NetLogo Simulation Engine–VERA Implementation}
VERA does not use all of NetLogo’s capabilities. Instead, it is structured as a virtual machine defining a limited subset of NetLogo methods. These include startup and setup methods, as well as methods for each of the Components and Relationships. Startup initializes parameter values obtained from the user’s Conceptual Model for each NetLogo breed. Setup initializes the individual agent parameters with additional information such as randomizing agent X-Y locations. For example, if the user is running a simulation with domestic sheep, setup could initialize a variable to store the breed’s expected lifespan of 251 ticks and then startup would place this value into the number of agents specified by the starting population value in the user specified values in the Conceptual Model. 

\begin{figure}[h]
  \centering
  \includegraphics[width=1.0\textwidth]{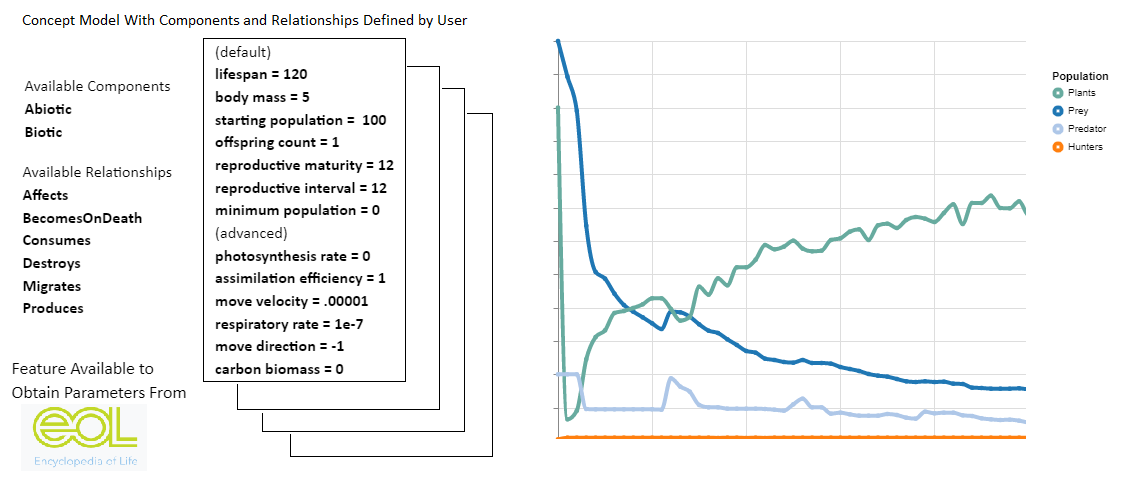}
  \caption{Front-End Data with Model and Simulation}
  \label{fig:fig7}
\end{figure}

As shown in Figure \ref{fig:fig7}, the NetLogo engine uses a customized method based on the user’s conceptual model and produces population levels for each tick. In addition to the properties based on user parameters, there are additional properties set for each of the agents as required by the NetLogo engine such as their X-Y placement within a 32x32 grid and an identifying number. Runtime processing is then controlled by the simulation engine’s main loop. 

Each Component and Relationship has a set of methods called by NetLogo's “go” loop to be run for each tick time. For example, biotic Components have predetermined methods for reproduction, limiting lifespan and processing biochemical resources to make sure that the Component has enough energy to live. An example Consumes Relationship causes the wolf agents to update their energy levels based on their Consumption Rate and Interaction Probability parameters. Upon completion of each tick, the resulting population levels are reported back to the user interface. The interface then displays an update to the line graph showing Component population levels for species. 

\begin{figure}[h]
  \centering
  \includegraphics[width=1.0\textwidth]{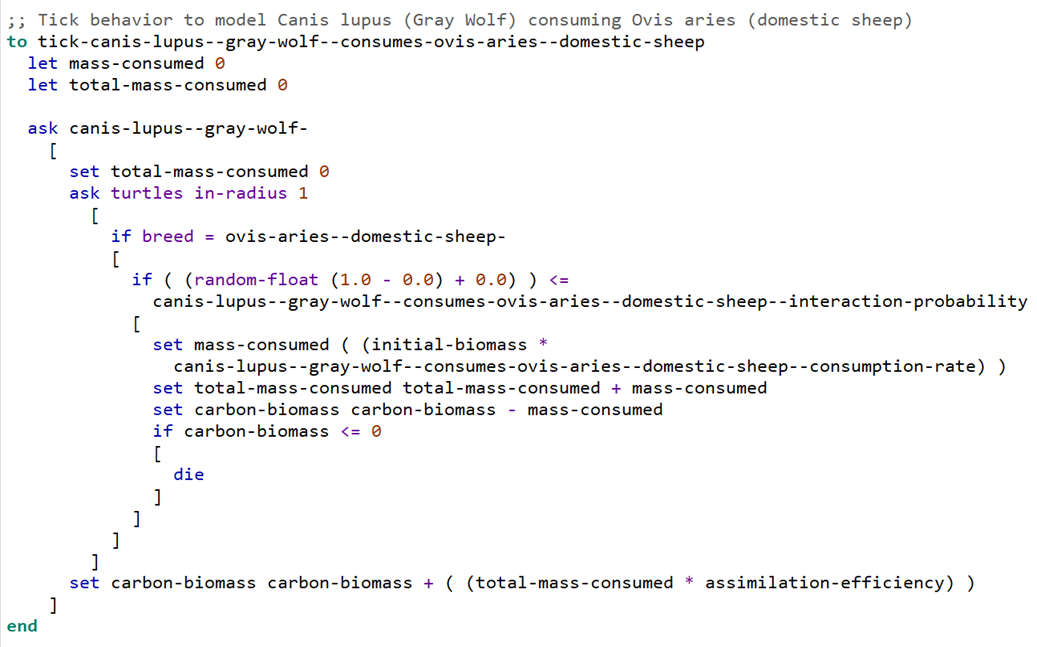}
  \caption{Automatically Generated VERA Procedure to Handle Interaction "Wolf Consumes Sheep"}
  \label{fig:fig8}
\end{figure}

In Figure \ref{fig:fig8}, VERA-generated NetLogo code is shown. The purpose of the procedure is to calculate the potential consumption of sheep by wolves. A random number is compared to the Interaction Probability. If the random number is less than the Interaction Probability the method continues, and the consuming Component increases its Carbon Biomass at a cost to the consumed Component. If the consumed Component has a biomass of zero or less, it is removed from the simulation. 

\subsection{Compilation Process}
The VERA compiler is responsible for processing the visual language model and producing NetLogo code compatible with the virtual machine API. Each Component and Relationship in the conceptual model is represented in XML for later processing and storage. The XML is later used by the VERA compiler as input to produce a customized NetLogo method. Table \ref{tab:t6biotic_compilation} describes how some of the VERA parameters are realized as compiled NetLogo methods. Table \ref{tab:relationship_compilation} does the same for the Relationship parameters. 

\begin{table}[h!]
\centering
\caption{Compilation Strategy for Biotic Parameters}
\label{tab:t6biotic_compilation}
\begin{tabular}{lp{10cm}}
\hline
\textbf{Parameter} & \textbf{Description} \\
\hline
Lifespan & Removes an agent when it reaches its age limit \\
Minimum Population & Generates agents if the population is small enough. This is used to simplify a system, for example, to enable a food source to be available without specifying its underlying life cycles \\
Biomass & Calculates incoming biochemical energy through photosynthesis-rate and outgoing energy reserves through respiratory-rate \\
Reproduction & Creates an additional instance of the agent in the simulation using the reproductive-maturity, offspring-count, and reproductive-maturity parameters \\
Movement & Places of the agent within the simulation using the move-direction and the move-velocity parameters \\
\hline
\end{tabular}
\end{table}

\section{VERA's Classroom Implementation}
VERA has been implemented in a variety of educational settings, including use by undergraduate and REU students at the University of Tennessee at Chattanooga, summer interns at the National Museum of Natural History, and self-directed learners on the internet \cite{44_an2022effects, 39_agarwal2018from, 40_an2019learning, 41_an2020scientific, 42_an2021recognizing, 43_an2022contextualizing}. At the Georgia Institute of Technology, VERA has been incorporated into several undergraduate courses, such as Introduction to Biology, General Ecology, Ecology Laboratory, and Biostatistics, as well as graduate-level courses in Computational Creativity and Cognitive Science for computer science students.

\begin{table}[h!]
\centering
\caption{Compilation Strategy for Relationships}
\label{tab:relationship_compilation}
\begin{tabular}{lp{10cm}}
\hline
\textbf{Parameter} & \textbf{Description of Parameter} \\
\hline
Consumes & Considers nearby agents that have been marked for possible consumption. Consumption is limited by interactive probability. The biotic being consumed transfers a portion of its carbon-biomass to the consumer. Carbon-biomass is consumed as energy by the biotic Components. An agent that has all remaining carbon-biomass reduced to 0 is removed from the simulation \\
Destroys & Removes the targeted biotic based on an interactive-probability and a consumption-rate. If the consumption rate is less than 1, reduction of the agent’s biomass is performed to simulate partial destruction \\
Affects & Causes an increase to the growth-rate of the targeted Component limited by interactive-probability \\
Produces & Causes an increase in the number of agents in the simulation based on a production-rate parameter \\
\hline
\end{tabular}
\end{table}

Empirical studies show that VERA improves students’ conceptual understanding of ecology \cite{40_an2019learning}. For example, in Fall 2018, VERA was introduced into a General Ecology course at the Georgia Institute of Technology. Among the enrolled students (N = 91), 52 completed both a pre-test and post-test assessing biology-related concepts. The difference between the two (t = 2.171) was statistically significant, indicating higher post-test scores and demonstrating gains in conceptual knowledge following interaction with VERA.

Since its public release in Fall 2018, VERA has been freely available and globally accessible through multiple platforms, including the Smithsonian Institution’s eol.org. Over this period, several thousand learners of diverse and largely unknown demographics have used the system. Of these, 315 learners constructed and/or simulated a total of 822 models in sufficient detail to allow analysis of their modeling behavior. Learners who completed the full exploratory cycle—model construction, parameterization, and simulation—demonstrated the scientific reasoning patterns described in the literature \cite{20_clement2008creative, 24_schwarz2009learning, 44_an2022effects, 40_an2019learning, 43_an2022contextualizing, 46_nersessian2008creating}.
\section{Related Work}
Work related to VERA has been conducted in the areas of education, modeling, automated tools and compilation. 

\subsection{Education}

The interactive approach of problem solving that VERA facilitates through cognitive assistance implements constructionism \cite{1_papert1991situating}, which itself is based on the constructivist theory of education. Constructivist theory \cite{47_piaget1964cognitive} presents the idea that learning growth is accomplished by having students actively engage with the material being learned and with fellow students. Learners add information into their current understanding and build upon their previous experiences. It directs instructors to focus on facilitating experiences with problem solving and critical thinking based on student prior knowledge rather than traditional direct instruction. Constructionism builds on constructivism and points out that best practices incorporate technology into a project-based approach to education. Emphasis is placed on the project being relevant to students’ personal experiences and the use of a hands-on design to facilitate learning. 

Betty’s Brain \cite{48_andres2019affect} is an implementation of constructionism comparable to VERA’s. Betty’s Brain challenges students to design a conceptual map as they teach a virtual character named Betty about an ecosystem. It has been shown to facilitate learning in 6th grade students and shows connections in emotional state with corresponding academic achievement. Emotional states such as boredom, confusion, engagement, delight, frustration and surprise were evaluated during a student experiment. They found evidence that students who had low pretest scores on the subject were more likely to be bored. In Betty's Brain, learners can easily draw a qualitative diagram with some labels indicating positive relationships among variables, such as increase, create, raise, or come from, and negative relationships, such as blocks, between variables. Additionally, learners can ask questions about what happens to one variable when another variable's value is changed. 

The Beyond Black Boxes (BBB) initiative \cite{49_resnick2000beyond} implemented a constructionist approach to education where students created their own designs and tools for scientific investigation. The study showed that the practice of instrument building can enable developing scientists to gain a deeper understanding of the underlying processes of the world. Unless educators are careful, students may become disconnected as they measure the world with instruments that they do not understand. Because computational technology can be especially opaque to the underlying processes, projects like BBB enable users to have accessibility to the tools of analysis to customize to their own needs for scientific tools. 

\subsection{Modeling}

VERA promotes learning by enabling students to create and refine conceptual models and then validate them via simulation. The experiments so devised can be adjusted according to problem evolution and increasing student understanding. Here are comparable works that have used similar modeling approaches. 

Hestenes gives an interesting explanation and outlines a well thought out vocabulary of the philosophical development of modeling with intelligence. His paper gives evidence for a modeling theory of mind to guide research in cognitive science \cite{50_hestenes2015conceptual}. The idea is presented that mathematical rules are for theorists; measurement standards are for experimentalists; and probability theory is for those who study phenomena. VERA is aligned with the ideas expressed in this work in that it helps a user express a mental model as a conceptual model. 

According to Koperski \cite{51_koperski2020models}, a model is a representation of some object, behavior, or system that one wants to understand. It creates an analogy between the model and its subject, implying a similarity in their observable properties. Models lead to learning in three stages: denotation, demonstration and interpretation. Denotation establishes a representation relation between the model and the target. Demonstration investigates the features of the model to demonstrate theoretical claims about its internal constitution or mechanism. Interpretation then converts knowledge about the model into knowledge about the target. VERA enables a user to express a denotation in a Conceptual Model. It then facilitates demonstration via an agent-based simulation engine. 

\subsection{Tools}

VERA is a cognitive assistance tool that enables users to construct conceptual models of ecosystems. There are comparable cognitive tools that can be considered when understanding VERA’s approach. 

Prometheus \cite{15_bridewell2006supporting} implements a user-accessible iterative cycle of development through a process model and a simulation model in much the same way that VERA uses a conceptual model to generate a simulation. However, Prometheus requires the user to select rates of change for Components to be calculated using differential equations rather than via the agent-based simulation that VERA uses. The Prometheus platform is designed to run on a user’s machine rather than the VERA web-accessed server-based platform. The Prometheus development environment enables a user to build a process model and produce a graph of simulated trajectories in much the same way that VERA uses a conceptual model and a simulation engine. 

Another approach comparable to VERA’s is an object-oriented language extension called OOCSMP \cite{52_alfonseca1998education,53_deLara1999simulating}, that enables users to generate code for a variety of targets, including C++, Java and HTML. The system uses automated simulation creation software as a strategy for setting up curricula for ecology courses. OOCSMP uses equations designed to specifically model the interactions between predator and prey populations to show the dynamics of population changes over time. It uses an object-oriented continuous simulation language to express the models. 

In a discussion of an overview of the general state of ABM \cite{54_abar2017agent} presents a strategy of comparing ABM systems in terms of ease of use, scalability and computational strength. The authors' recommendations place NetLogo on the top level with respect to ease of use and a somewhat high level of strength and scalability. VERA compares well as far as ease of use is concerned as it is accessible via the web and does not require programming. 

An additional strategy for classifications of ABMs is asserted by \cite{55_taillandier2019building} in their paper describing the GAMA (GIS Agent-based Modeling Architecture) platform. GAMA is like VERA in that it provides access to complex modeling systems without the need for extensive computer science knowledge. In their presentation of GAMA 1.8, the authors offer a three-tiered approach to classifying visually based programming languages for agent-based simulations: 1) powerful and less accessible platforms that are useful for experienced programmers to use within existing common tool sets such as Python and C++; 2) platforms that are designed to increase accessibility to those with a wider range of technical skills such as GAMA and NetLogo; and 3) graphically based platforms that enable the user to design models with tools such as StarLogo \cite{56_resnick1996starlogo} and AgentSheets \cite{57_repenning1995agentsheets}. 

Repast Simphony \cite{58_north2013complex} enables programmers to create settings for complex behaviors in order to model systems relying on agent choice. Unlike VERA’s target users (students), Repast Simphony is a system that increases access for computer programmers to simulate complex agent behavior depending upon the needs of the experiment. It focuses on the idea of making model software more modular. VERA’s compiler architecture follows this philosophy in that, although its structure is set up to access NetLogo, it is designed to be programmatically interchangeable with other engines. 

Jaiswal \cite{59_jaiswal2019comparative} compared conceptual models and physical models used to analyze rainfall runoff and the structure used to manage it. For example, in a watershed, both models analyze the process of water evaporating, becoming precipitation and running off into a basin. In the conceptual hydrological models, simplified mathematics describes the different components that store and distribute water. The physical models use a more complex approach that considers mass, energy and momentum and require a large amount of data from the environment such as from soil, land use and climate. The information it produces can be used to design systems to create cost-effective approaches for water management in developing countries. The article analyzes the strengths and weaknesses of the different kinds of models and gives real examples of the results from the two categories. The scientific process expressed in this article (topological water flow from rainfall) uses several different models from the two categories. The established logical process of beginning with conceptual models and then moving to physical models is done separately. VERA, in contrast, provides the user a higher level of abstraction through a conceptual model and then uses empirical parameters to populate an agent-based simulation. 

EcoSim \cite{60_christensen2004ecopath} is an agent-based ecology simulation tool focused on modeling predator-prey interactions. Unlike VERA, it supports speciation, genomic data, evolving relationships, sensory inputs, and internal states (e.g., fear, hunger). Moreover, it supports large-scale interactions lasting tens of thousands of time steps and up to a billion agents. Ecopath builds on EcoSim as an ecological modeling software suite that focuses on marine science with currently more than 8,000 users in over 170 different countries. The Ecopath part of the suite provides a static, mass-balanced snapshot of a system. EcoSim adds a dynamic simulation module for policy exploration. A third component, Ecospace, offers a spatial and temporally dynamic module primarily designed for exploring impact and placement of protected areas. Interestingly, the suite offers variable-speed splitting, which enables efficient modeling of the dynamics of both “fast” (e.g. phytoplankton) and “slow” (e.g. whales) biotic species. Like VERA, the suite allows users to load time series reference data. 

\subsection{Compilation}

VERA compiles a user-drawn model of an ecological system into a NetLogo simulation program. The compiler is syntax-directed in that each of its syntactic units (Components and Relationships) is separately compiled into a NetLogo method. The methods are combined with pre-built infrastructure methods before being sent to a NetLogo server. There are some other tools that take comparable approaches in their compilation process. 

OpenABL \cite{61_cosenza2021easy} is a programming environment designed to simulate agent-based simulations and state machine approaches. The environment focuses on performance, programmability, portability and reproducibility. The OpenABL compilation process divides the work into parsing and checking, intermediate representation and high-level code transformations. There are multiple backend target choices. The parsing and checking portion prepares the compilation process by creating an abstract syntax tree that validates language semantics ensuring proper use of language constructs such as functions, loops and agents. An intermediate representation sets up a node-based structure. High-level code transformations then optimize the intermediate representation. The backend supports Flame, FlameGPU, Mason and D-Mason. Flame is a simulation framework that is based on state machines, and FlameGPU extends this by making the software able to run on GPU hardware. Mason is a Java library for agent-based modeling with D-Mason extending it to enable the software to distribute processing across multiple-core systems. 

MADRaSA (Multi Agent Driving Simulator) \cite{62_santara2021madras} is an open-source simulator for motion planning in autonomous vehicles. It uses Markov decision processes, Markov games, reinforcement learning and episodic learning to produce motion planners. The simulator can have some cars that are using automated learning systems and some that are not learning to create a more realistic model. The system contains six predetermined types of cars to create a challenging environment for the controller to navigate. VERA uses a similar idea with scaffolding predetermined methods that make biotic Components and Relationships functional in the agent-based system. 

Kary \cite{63_kary1986trc} describes a software tool called TRC (Translate Rules to C). It enables the user to automatically generate C code from a set of rules. The system contains a translator of a rule-based language into a procedural language that improves speed and convenience. TRC describes expert problem-solving systems in terms of situation-action rules that take advantage of known system structures and information that varies in each instance. They refer to these as “Long Term Memory” and the varying properties as “Short Term Memories”. VERA is comparable to this approach in that it has a set of predetermined methods that dictate typical biological processes. These processes determine an organism's survival in an ecological system. The specific parameters of each organism and their relationship to other Components are customized by the user. 

Finally, the Modeling, Simulation and Verification Language (MSVL) \cite{64_yang2018compiler} is a programming language designed to support non-deterministic and concurrent programming: that is, for the construction of programs modeling complex, real-world situations. MC is a compiler for MSVL. MSVL is based on temporal logic: the formal description of situations involving time. Because it is formal, programs can be verified to have certain properties, like that which is done by model checkers. Moreover, MSVL’s internal representation enables reasoning over all possible program executions, which the authors call modeling. MC is like VERA in that it is agent-based but goes well beyond it with respect to formality. Its domain of application is primarily state-based concurrency, whereas VERA is more stochastic. 
\section{Discussion}

\subsection{Current Deployment Status}

\begin{figure}[h]
  \centering
  \includegraphics[width=1.0\textwidth]{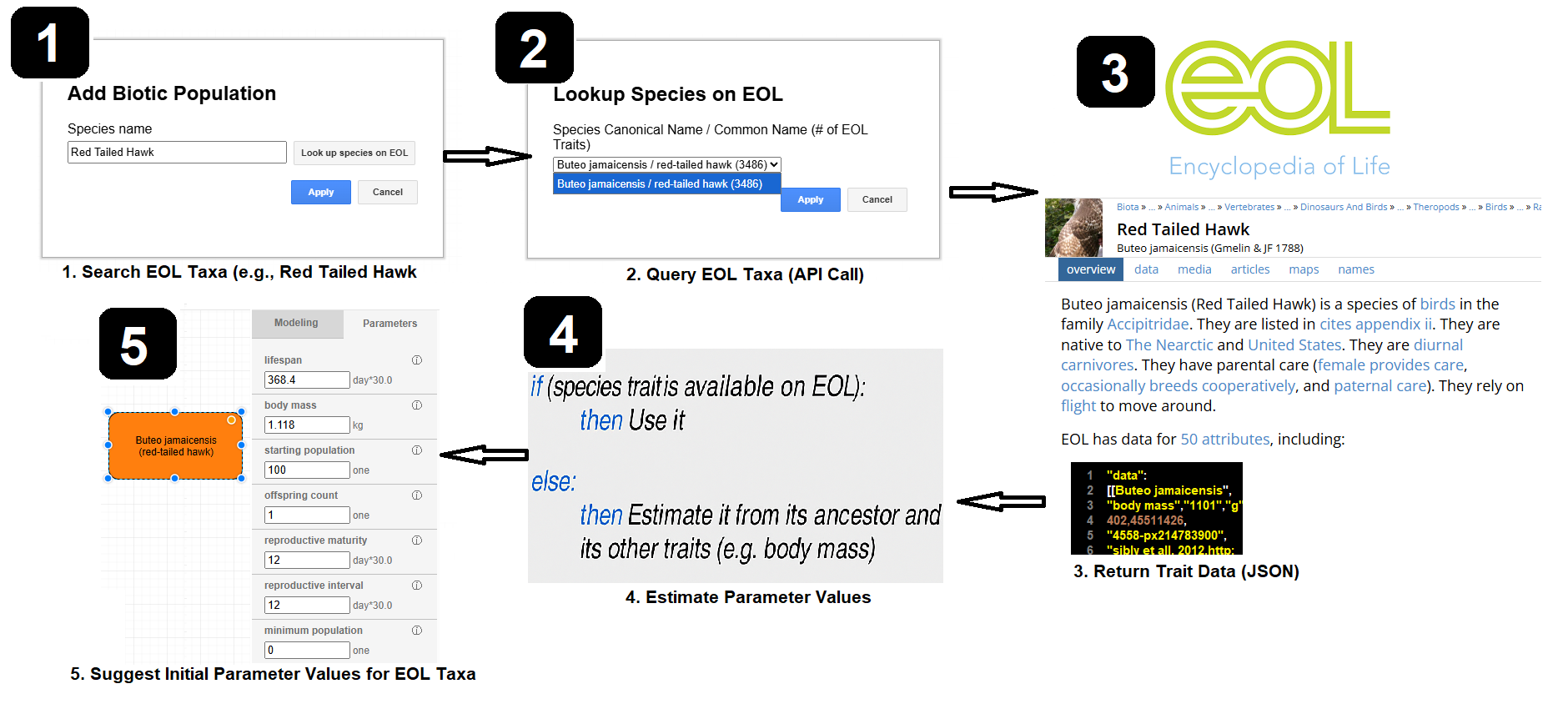}
  \caption{EOL Trait Bank Data to Set Simulation Values}
  \label{fig:fig9}
\end{figure}

VERA is hosted on the cloud using Amazon Web Services (AWS). The application is packaged and deployed on AWS Elastic Beanstalk which runs on an Amazon Elastic Compute Cloud (EC2) instance, a form of Infrastructure-as-a-Service  that provides a scalable compute platform for developing and deploying applications. Extensions to VERA, including VERA Analytics, host Python code used for conducting A/B experiment analysis and other analytics capabilities \cite{65_hornback2023scalable, hornback2023conductingabexperimentsscalable} and the use of EC2 instances as the underlying virtual machines. The EC2 platform enables VERA to auto scale capacity depending on user needs, providing a seamless mechanism that can handle increased load capacity, such as when VERA is being used to run classroom experiments. All EC2 instances run on Amazon Linux. 

The backend database used to store VERA data is also hosted on AWS using its Relation Database Service (RDS). RDS provides a solution to VERA’s data needs including scaling, high availability and backups that are managed via AWS and ensure long-term data integrity. 

Source control of VERA is maintained via GitHub. All development and testing are conducted locally via the development team and only sent to final production after meeting internally developed software engineering protocols. 

\subsection{Limitations}

\begin{figure}[h]
  \centering
  \includegraphics[width=1.0\textwidth]{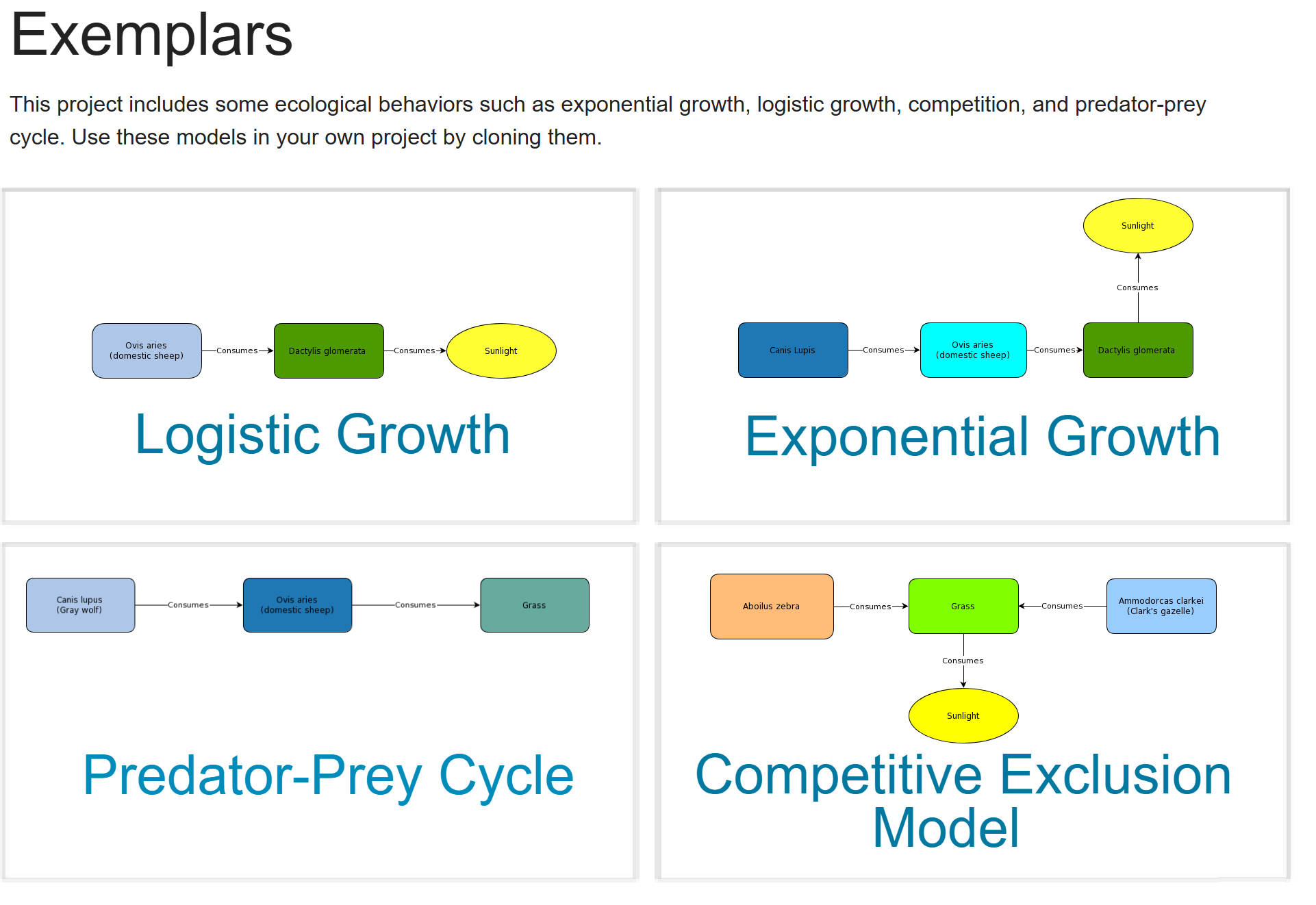}
  \caption{Exemplars to be Copied and Customized}
  \label{fig:fig10}
\end{figure}

As a higher level abstraction of the underlying agent based simulation, experimental success has been limited to implementations where the instructor has carefully crafted the majority of parameters and directed the student user to focus on some or a single parameter to make an analysis of the changes to the overall ecological emergence \cite{bunin_habitats}. Although this results in some useful results for specific lessons concepts such as understanding the predator-prey model, it disables the end user from observing the underlying process and therefore makes it impossible for them to create their own ecological systems without first building off of an instructors design or one of the built in examples in the web app. Future software may create additional educational opportunities by giving instructors features that can enable/disable access to the underlying agents and their interactions.   

As the current version of VERA is implemented with NetLogo, VERA is limited to solving the types of problems that are suitable using NetLogo. For example, the creators of NetLogo point out that tracking all the molecules in a system could lead to accurate results but is not practical using their tool. In systems where the agents are homogeneous in nature, there may be other approaches that use summative equations to express the models. Agent-Based Simulations within NetLogo work well for ecological systems that consider a limited number of component types where the individual agents contain heterogeneous state such as locations and biochemical amounts, but it would be impractical to calculate every possible state of the system \cite{66_wilensky2015introduction}. 

Other ABS frameworks can significantly increase the scale of model complexity. Supercomputers implementing the FLAME framework have been shown to be useful in biological science in predicting behavior on the cellular level, such as the movements of molecules within the cytoplasm of simulated living cells \cite{67_holcombe2022agent}. This is beyond the scope of current VERA design, but the VERA conceptual model approach could potentially be applied to emerging ABS technologies. 

VERA simulations are simplified to make the computational processes manageable. The simulated environment is limited in scope and size. There are a limited number of parameters associated with the individual Components and the values of these parameters are limited in precision. The order of agent processing is fixed and sequential unlike natural processes that occur simultaneously. These constraints can combine to reduce the accuracy and precision of results when compared to more elaborate scientific simulators. 

The agent environment of VERA is calculated in a two-dimensional world on a 32x32 grid. Moreover, when agents move off an edge, they appear on the opposite edge. Agent mass is not considered when computing movement. Within the same grid cell, agents are not considered to be above or below each other. 

The state of the agents in VERA is limited in complexity. To reduce load in server processing, VERA simulations have a maximum of 25,000 agents and methods that create additional agents, such as reproduction, will halt if this ceiling is reached. Agents are given a maximum age and are removed from the simulation when this number is reached. These kinds of constraints are part of the nature of simulations and may lead to deviation from real systems. 

\subsection{Encyclopedia of Life}

VERA retrieves large-scale domain knowledge from the Smithsonian Institution's Encyclopedia of Life (EOL). EOL is a highly curated and comprehensive database of species data, featuring nearly two million species and over eleven million trait data records in the biological domain (eol.org) \cite{33_parr2014encyclopedia}. The data from EOL is used to initialize parameter values for each species. 

EOL provides a wealth of trait data, aggregating records from various sources and studies conducted under diverse conditions. To facilitate large-scale queries, EOL offers API services that provide on-demand JSON output, presenting a concise and informative summary of knowledge related to EOL taxa, ecological interactions, and organism attributes. This makes it easier for researchers to access and use the vast collection of data available in EOL. 

VERA provides the contextualized biotic trait (parameter) data from EOL via the “Lookup EOL” feature for suggesting initial parameter values. Figure \ref{fig:fig9} illustrates the five steps involved in adding a biotic Component through this feature. First, the student queries a species name (using either a scientific or a common name). Second, VERA returns a list of species names that match the input via the EOL Search API. Third, the student selects one species from the list, and VERA calls the EOL TraitBank API to retrieve specific traits of the species. Fourth, VERA uses the retrieved traits to estimate the simulation parameters. Lastly, the simulation can be run with the preset values, or the learner can modify them as desired. 

\subsection{Exemplars}

Exemplar models are available to VERA users. As seen in Figure \ref{fig:fig10}, there are four stereotypical ecological behaviors available. In a model exhibiting logistic growth, a population initially undergoes an exponential increase in population until a resource it depends on is exhausted and the population reaches the largest value its ecology supports. Exponential growth provides a model where the biotic population continues to increase without bound because the ecosystem provides sufficient resources, lack of competition or the absence of predators. The predator–prey exemplar shows a dynamic situation where the populations vary cyclically. In the competitive exclusion exemplar, the simulation can explore the results of population changes due to multiple biotic types relying on the same resources. The exemplars give the user a jump start on creating a model that can then be refined.

\subsection{Systems Thinking}

The VERA tool is primarily concerned with ecological modeling. Its compiler specifically knows about Component and Relationship parameters such as lifespan and interaction probability. However, in principle, VERA is much more general than this. In fact, the VERA models shown in this paper are instances of the VERA ecology meta model (MM), which serves as a partial definition of the VERA visual modeling language. Notice how terms like "Biotic" and "Consumes" are directly apparent in the MM. 

\begin{figure}[h]
  \centering
  \includegraphics[scale=0.75]{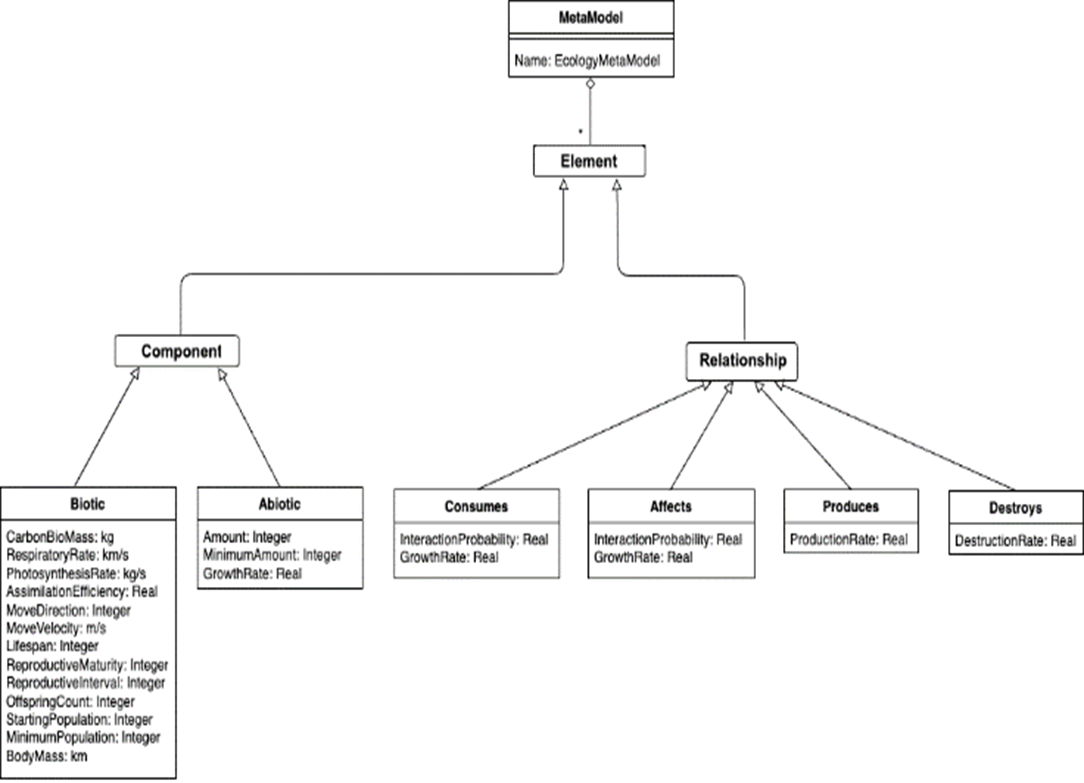}
  \caption{Vera Ecological Meta-Model}
  \label{fig:fig11}
\end{figure}

Figure \ref{fig:fig11} is expressed as a UML (Unified Modeling Language) static model diagram \cite{68_omg_uml}. In these diagrams, rectangles correspond to types of objects. Each rectangle can be partitioned into horizontal regions. Here two regions are used: the top one giving a name to the object type, and the lower one listing attributes of the instances of that type. Lines between rectangles denote associations between the types. Two categories of associations are of particular importance. An association ending with a triangle indicates subtyping. That is, the type at the unadorned end is a subtype of the type at the adorned end, thereby inheriting all of its attributes. For example, sperm whale is a subtype of cetacean. 

If instead of a triangle, the adornment is a diamond, then the association indicates that instances of the type at the unadorned end are part of the type at the adorned end. This type of association is called an aggregation. For example, a modeled ecology is composed of its biotic Components and their Relationships. Further, if the diamond is filled in, the association is called a composition, indicating that the source end type represents an inherent and inseparable element of the target end type. For example, a human microbiome is an essential part of a human. 

Finally, if the line between two types does not sport an adornment, then the corresponding association is more general. Typically, these lines will be annotated with a name, giving some interpretation of the underlying association. VERA Relationships, such as Consumes, indicate the two biotic Components each exist separately, but nevertheless are essentially related parts of the model. 

\begin{figure}[h]
  \centering
  \includegraphics[scale=0.75]{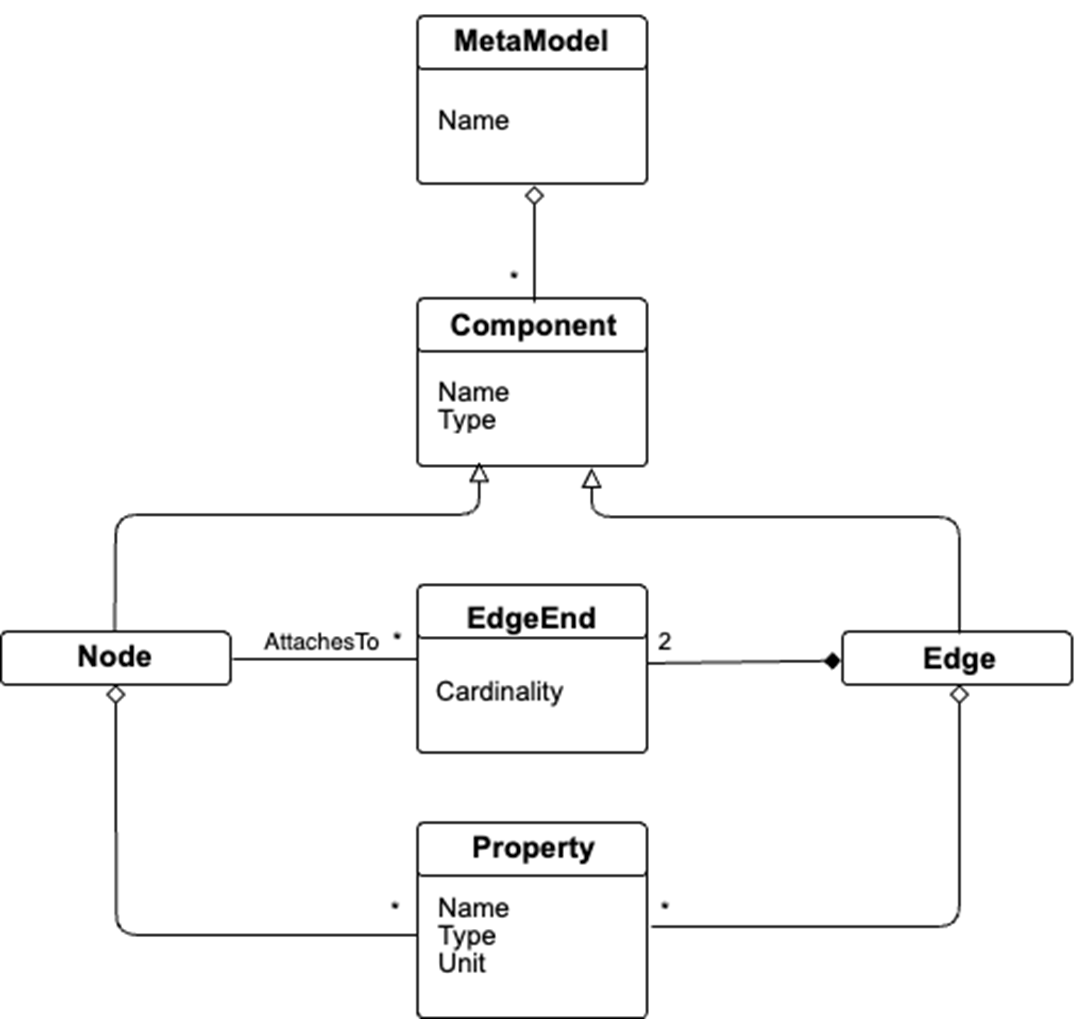}
  \caption{Vera Meta-Meta-Model}
  \label{fig:fig12}
\end{figure}

The VERA MM generalization can be taken one step further. Figure \ref{fig:fig12} shows the VERA meta-meta-model (MMM), suggesting how VERA can be generalized across other domains. In fact, instances of the VERA MMM have been used to create prototype meta-models for the domains of personal finance and epidemiology \cite{69_broniec2021guiding}. 

Part of the VERA vision is to automate the generalization enabled by the VERA MMM. Imagine that an instance of the VERA MMM was built for the domain of astronomy. Nodes might represent celestial objects such as planets, satellites, asteroids, and the sun. The major association would denote the Attracts Relationship (gravity). The generalization would then use the astronomy Meta-Model to generate an astronomy-specific compiler for building solar system simulations. 

\subsection{Ask Jill}

A major concern in deploying VERA in educational settings is usability and learnability of the interactive modeling environment. To enhance VERA's usability and learnability, VERA provides users with access to both tutorials in the form of videos and detailed user guides in the form of textual documents. However, it was found that not all users took advantage of these materials. Thus, a conversational agent called AskJill was embedded in VERA to answer users' questions about what VERA does, how it works, and how a user may use it to address a problem \cite{70_goel2021explanation}. AskJill builds on our earlier work on Jill Watson, a virtual teaching assistant for answering students' questions on discussion forums of online classes \cite{71_goel2018jill}. 

\subsection{Personalization}

Personalization aims to provide individualized learning experiences that help each learner to achieve their maximum potential. The VERA project affords investigation of personalization of learning in the context of ill-defined problems and self-directed learning. 

This effort is developing machine learning techniques, such as dimensionality reduction, hierarchical clustering, and Markov chain modeling, for analyzing the behaviors of self-directed learners addressing ill-defined problems \cite{41_an2020scientific}. This analysis has led to the identification of multiple types of modeling behaviors such as Construction, Observation, and Exploration. The project is also developing a suite of interactive coaches that can nudge individual learners towards more productive behavior based on an analysis of their previous actions. For example, VERA detects that if a learner is repeatedly engaged in Construction behavior, a coach within VERA may nudge the learner towards Exploration. 
\section{Conclusion}

VERA is an interactive learning environment used to support the teaching of introductory ecology in college courses. It enables students to develop conceptual models of an ecosystem and then compare them against archetypical behaviors such as predator–prey cycles. Model validation is done by an independent agent-based simulation tool called NetLogo. The major contributions of VERA include its effective use of conceptual modeling, its model-validation via simulation, and most importantly, its automatic translation of the former to the latter. In particular, no coding is required of students. Classroom experiments using VERA have demonstrated improved learning. VERA is publicly available and has recently been exported to classrooms at other institutions as well as to the general public via the Smithsonian Institution’s EOL website.

\section*{Declaration of Competing Interest}

The authors do not have any conflict of interest with the research reported here. This research has been supported by various grants over the years including US NSF Big Data Spokes Grant \#1636848 and US NSF National AI Research Institutes Grants \#2112532 and \#2247790. The use of Encyclopedia of Life (EOL; eol.org) in the VERA learning environment is supported by Smithsonian Institution’s Encyclopedia of Life Project. GloBI and NetLogo, used in VERA, are available as open source services from https://www.globalbioticinteractions.org/ and https://ccl.northwestern.edu/netlogo/, respectively. Georgia Tech owns the copyright to VERA. 

\section*{Acknowledgments}
This research has been supported by NSF Grants \#1636848, \#2112532 and \#2247790. We would also like to thank Robert Bates, Jennifer Hammock, David Joyner, Greg Newman, Emily Weigel and student members of the Design Intelligence Laboratory at Georgia Tech for their contributions to this work.

\section*{Data Availability}
The VERA interactive learning environment is accessible publicly at vera.cc.gatech.edu. This research was conducted in accordance with protocols approved by the Georgia Tech IRB office that assure confidentiality of raw data. Thus, we are unable to share any data on human subjects. 

\clearpage


\end{document}